\documentclass[epj]{svjour}

\usepackage{amssymb}
\usepackage{amsmath}
\usepackage{graphicx}
\begin{document}

\title{Spherical Collapse of a Unified Dark Fluid with Constant Adiabatic Sound Speed}

\author{Lixin Xu \thanks{lxxu@dlut.edu.cn}}

\institute{Institute of Theoretical Physics, School of Physics \&
Optoelectronic Technology, Dalian University of Technology, Dalian,
116024, P. R. China\\
College of Advanced Science \& Technology, Dalian
University of Technology, Dalian, 116024, P. R. China}

\abstract{
In this paper, we test the spherical collapse of a unified dark fluid (UDF) which has constant adiabatic sound speed. By choosing the different values of model parameters $B_s$ and $\alpha$, we show the nonlinear collapse for UDF and baryons which are considered for their formation of the large scale structure of our Universe. The analyzed results show that larger values of $\alpha$ and $B_s$ make the structure formation faster and earlier.}

%\pacs{98.80.-k, 95.35.+d, 95.36.+x}

%\keywords{Unified dark fluid, spherical collapse} 

\maketitle

\section{Introduction}

A unified dark fluid (UDF) model \cite{ref:darkdeneracy,ref:Bruni,ref:darkdeneracyxu,GCG,GCG-action,GCGpapers,GCGdecomp,GCGxu,ref:xuNUDF,ref:alphcdm,ref:cs0,ref:csvar}, which unifies cold dark matter and dark energy as a unique component, was considered as an alternative model to explain the late time accelerated expansion of our Universe \cite{ref:Riess98,ref:Perlmuter99}. Usually the UDF behaves like the cold dark matter at the early epoch and like the dark energy at the late time. On the background level, the UDF model can mimic the $\Lambda$CDM model very well. And the cosmic observations from the type Ia supernova (SN), the Baryon acoustic oscillation (BAO), and the full information of the cosmic background radiation (CMB) can not distinguish the $\Lambda$CDM model from the UDF model, for example please see \cite{ref:darkdeneracyxu,GCGxu,ref:xuNUDF}. 

But can the UDF form the observed large scale structure of our Universe? The primordial quantum perturbations seed the large scale structure of our Universe. And the small perturbations grow into the large scale structure via gravitational force due to the local over density contrast and gravitational instability. So for any realistic cosmological model, it is demanded to describe the whole background evolution history and the formation of the observed large scale structure. It is also the situation for the UDF models without exception. If a cosmological model cannot describe the observed large scale structure, it would be ruled out through a comparison of the theoretical calculation and the cosmic observations. When one investigates the evolutions of density perturbations of UDF, the non-linear stages of perturbations are not avoided. However, the study on the non-linear evolutions is expensive when one uses the hydrodynamical/$N$-body numerical simulation. Before the expensive investigation, we should test the viability for the formation of the large scale structure via a simple model or framework. Fortunately one can investigate the non-linear stages of collapse in the framework of spherical "top-hat" collapse and leave the cumbersome hydrodynamical/$N$-body numerical simulation alone. 

Actually, in the last few years, the generalized Chaplygin gas (gCg) model \cite{GCG,GCG-action,GCGpapers,GCGdecomp,GCGxu}, which is an extension of the standard Chaplygin gas model \cite{ref:Chaplygin}, as a famous UDF model was discussed extensively. It covers the range from the background evolution and the linear perturbations to the non-linear collapse \cite{ref:Fernandes2012,ref:Abramo2009} in the framework of spherical collapse model. However when one studies the non-linear perturbations, one has to overcome the so-called averaging problem \cite{ref:darkdeneracyxu,ref:stability}. The problem comes from the fact that $\langle p\rangle\neq p(\langle\rho\rangle)$ for the gCg model \cite{GCG}
\begin{equation}
p=-\frac{A}{\rho^\beta}
\end{equation}
where $A$ and $\beta$ are positive model parameters, one can check that 
\begin{equation}
\langle p\rangle=-\langle A/\rho^\beta\rangle\neq -A/\langle\rho\rangle^\beta=p(\langle\rho\rangle),
\end{equation}
in the case of $\beta\neq 0$. However, it is not the issue for a model which has a linear relation between $p$ and $\rho$, i.e. $p=\alpha\rho-A$. This model has a constant adiabatic sound speed and has been investigated extensively in the last few years \cite{ref:Bruni,ref:darkdeneracyxu,ref:alphcdm}. It would be better to investigate the non-linear perturbation evolution in this model due to the escape from the averaging problem. But whether can this UDF form the large structure of our Universe? And what are the non-linear perturbation evolutions for this UDF? What are the effects of the model parameter on the collapse? To answer these questions is the main motivation of this paper. But instead of using the hydrodynamical/$N$-body numerical simulation, we will check the possibility of the collapse or the large structure formation in the framework of spherical collapse for the UDF with a constant adiabatic sound speed. 

The paper is structured as follows. In section \ref{sec:review}, we give a brief review of the UDF with a constant adiabatic sound speed. We present the basic equations for spherical top-hat collapse of fluid in section \ref{sec:basic}. The main results are summarized in section \ref{sec:perevolution}. Section \ref{sec:conclusion} is the conclusion.

\section{A Brief Review of Unified Dark Fluid with Constant Adiabatic Sound Speed}\label{sec:review}

In this section, we will give a brief review of a UDF model which has a constant adiabatic sound speed (CASS) $c^2_s=\alpha$, for the details please see the Ref. \cite{ref:darkdeneracyxu}. From the definition of the adiabatic sound speed
\begin{equation}
c_s^2=\left(\frac{\partial p_d}{\partial \rho_d}\right)_s=\frac{d p_d}{d\rho_d}=\rho_d\frac{dw_d}{d\rho_d}+w_d=\alpha,
\end{equation}
after integration,  one obtains
\begin{equation}
w_d=\alpha-\frac{A}{\rho_d},
\end{equation}
where $A$ and $w_d$ are the integration constant and the equation of state (EoS) of the UDF respectively. In terms of $\alpha$, one has the pressure of the UDF $p_d=\alpha\rho_d-A$. The energy density and the EoS of this UDF can also be recast into the following forms
\begin{eqnarray}
\rho_d&=&\rho_{d0}\left\{(1-B_s)+B_s a^{-3(1+\alpha)}\right\},\label{eq:rhod}\\
w_d&=&\alpha-\frac{(1+\alpha)(1-B_s)}{(1-B_s)+B_s a^{-3(1+\alpha)}},
\end{eqnarray}
where $B_s$ in the range $0\le B_s\le 1$ and $\alpha$ in the range $0\le \alpha\le 1$ are the model parameters. When $B_s$ is zero, this UFD is a cosmological constant. If $\alpha=0$ is respected, this UDF is the standard $\Lambda$CDM model at the background level. And when $B_s=1$ and $\alpha=0$ are set simultaneously, the cold dark matter is arrived. In a spatially flat FRW Universe, the Friedmann equation is given as
%\begin{widetext}
\begin{eqnarray}
H^{2}=H^{2}_{0}\left\{(1-\Omega_{b}-\Omega_{r})\left[(1-B_{s})+B_{s}a^{-3(1+\alpha)}\right] \right.\\
\left.+\Omega_{b}a^{-3}+\Omega_{r}a^{-4}\right\},
\end{eqnarray}
%\end{widetext}
where $H$ is the Hubble parameter with its current value $H_{0}=100h\text{km s}^{-1}\text{Mpc}^{-1}$, and $\Omega_{i}$ ($i=b,r$) are dimensionless energy density parameters for baryon and radiation respectively. 

In Ref. \cite{ref:darkdeneracyxu}, this model was tested by using SN Ia, BAO and the full information from WMAP-7yr data sets, where the model parameter space was obtained. The result shows that at the background level the currently available cosmic observations cannot distinguish the UDF model from the $\Lambda$CDM model. It means that they have almost the same background evolution history. However the dynamic evolutions would be strongly different even they have the same background evolution. One should use the large scale structure information to test the viability of UDF models as additions. Before doing that, at first we should check the possible collapse under this model. In the following section, we will use the spherical collapse model to investigate the non-linear evolution of the UDF perturbations. 

\section{Basic Equations for Spherical Top-hat Collapse of Fluid} \label{sec:basic}

The spherical collapse (SC) describes the evolution of a spherically symmetric perturbation embedded in a homogenous static, expanding or collapsing background. In this paper, for simplicity, we consider the spherical top-hat collapse (STHC) model which is characterized by the perturbed region with constant density \cite{ref:Fernandes2012}. With this assumption in hand, the perturbation only depends on the time. And one doesn't need to consider the gradients inside the perturbation region. 

In the STHC model, the basic equations which depend on local quantities can be written as
\begin{eqnarray}
\dot{\rho}_c&=&-3 h(\rho_c+p_c),\\
 \frac{\ddot{r}}{r}&=&-\frac{4\pi G}{3}\sum_i(\rho_{c_i}+3p_{c_i}),
\end{eqnarray}
where $h=\dot{r}/r$ and $r$ are the local expansion rate and the local scale factor respectively. Here the perturbed quantities $\rho_c$ and $p_c$ are defined was
\begin{eqnarray}
\rho_c&=&\rho+\delta\rho,\\
p_c&=&p+\delta p,
\end{eqnarray}
where $\rho$ and $p$ are background quantities which respect to the background evolution equations
\begin{eqnarray}
\dot{\rho}&=&-3H(\rho+p),\\
 \frac{\ddot{a}}{a}&=&-\frac{4\pi G}{3}\sum_i(\rho_i+3p_i),
\end{eqnarray}
where $H=\dot{a}/a$ is the Hubble parameter which relates to local expansion rate in the STHC model by \cite{ref:Abramo2009}
\begin{equation}
h=H+\frac{\theta}{3a},
\end{equation}
where $\theta$ is the divergence of the peculiar velocity $\overrightarrow{v}$, i.e. $\theta\equiv\nabla\cdot \overrightarrow{v}$.

The governing equations which describe the dynamical evolution of density contrast $\delta_i=(\delta\rho/\rho)_i$ and $\theta$ are given in the following form \cite{ref:Fernandes2012,ref:Abramo2009}
\begin{eqnarray}
\dot{\delta}_i&=&-3H(c^2_{e_i}-w_i)\delta_i-[1+w_i+(1+c^2_{e_i})\delta_i]\frac{\theta}{a},\label{eq:deltat}\\
\dot{\theta}&=&-H\theta-\frac{\theta^2}{3a}-4\pi Ga\sum_i\rho_i\delta_i(1+3c^2_{e_i}),\label{eq:thetat}
\end{eqnarray}
where $c^2_{e_i}=(\delta p/\delta \rho)_i$ is the square of the effective sound speed of energy component $i$. One can recast the Eq. (\ref{eq:deltat}) and Eq. (\ref{eq:thetat}) into the equations which evolve with respect to the scale factor $a$
\begin{eqnarray}
\delta'_i&=&-\frac{3}{a}(c^2_{e_i}-w_i)\delta_i-[1+w_i+(1+c^2_{e_i})\delta_i]\frac{\theta}{a^2H},\label{eq:deltaa}\\
\theta'&=&-\frac{\theta}{a}-\frac{\theta^2}{3a^2H}-\frac{3H}{2}\sum_i\Omega_i\delta_i(1+3c^2_{e_i}),\label{eq:thetaa}
\end{eqnarray}
where we have used the definition $\Omega_i=8\pi G\rho_i/3H^2$.

Observing the above dynamical evolution equations, one can find that the important involved quantities are the EoS $w_c$ of the collapsing region and the effective sound speed $c^2_e$. One can read off the EoS $w_c$ from its definition \cite{ref:Fernandes2012}
\begin{equation}
w_c=\frac{p+\delta p}{\rho+\delta\rho}=\frac{w}{1+\delta}+c^2_e\frac{\delta}{1+\delta}.
\end{equation}
So, the most important quantity is the effective sound speed $c^2_e$. For the CASS model, the effective sound speed is given as
\begin{equation}
c^2_{e}=\frac{\delta p}{\delta\rho}=\frac{p_c-p}{\rho_c-\rho}.
\end{equation}
By using the relation $p=\alpha\rho-A$, one has
\begin{equation}
c^2_{e}=\frac{(\alpha\rho_c)-A-(\alpha\rho-A)}{\rho_c-\rho}=\alpha
\end{equation}
without surprise. 

\section{Perturbation Evolutions of Density Contrast} \label{sec:perevolution}

Now, one can investigate the evolutions of density contrast of different energy component numerically in a spatially flat FRW Universe, where the relevant cosmological model parameters are borrowed from the results obtained in Ref. \cite{ref:darkdeneracyxu}: $H_0=71.341 \text{km s}^{-1}\text{Mpc}^{-1}$, $\Omega_{d}=0.956$ and $\Omega_{b}=0.044$. Here we just consider the baryon and UDF due to the possible formation of the large scale structure. We solve the differential equations by using the software {\bf Mathematica} and setting the initial conditions (ICs) $\delta_{d}$ and $\delta_{b}$ at the redshift $z=1000$ which is the same redshift used in Ref. \cite{ref:Fernandes2012}. 

\begin{center}
\begin{figure}[htb]
\includegraphics[width=8cm]{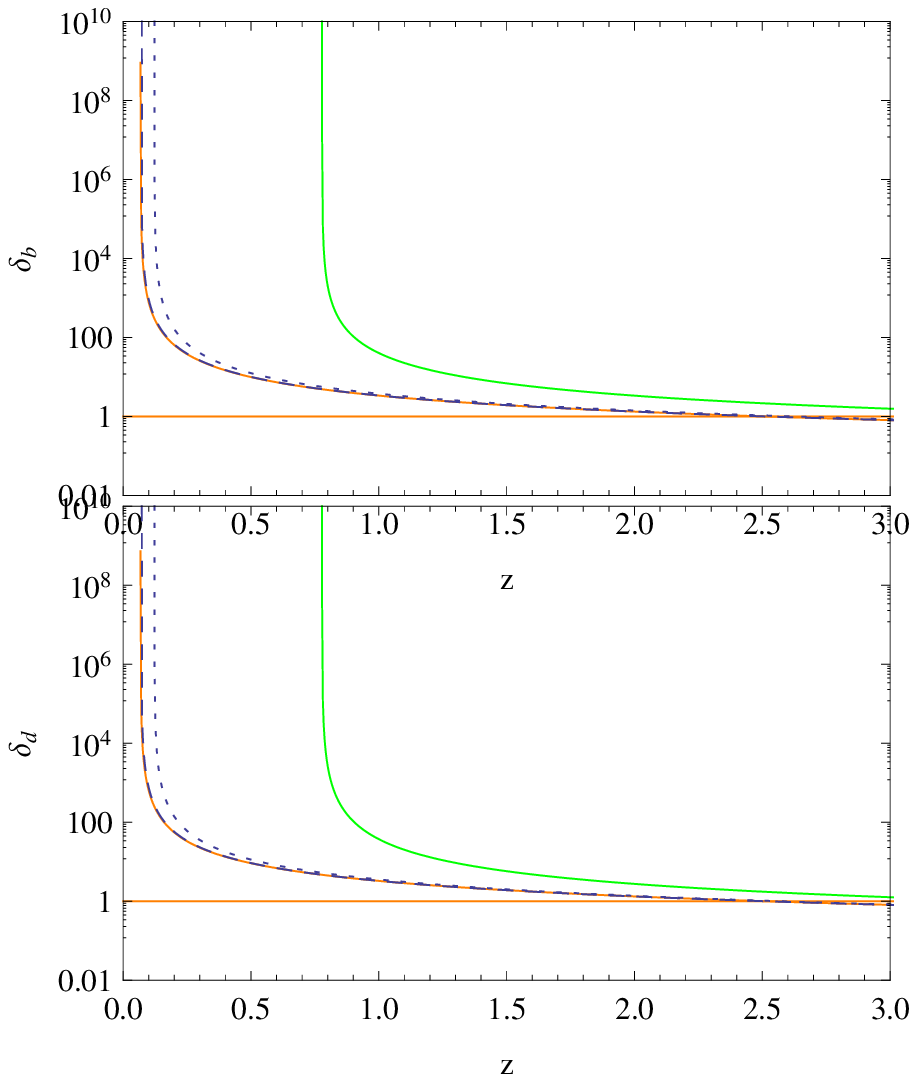}
\caption{The evolutions of density perturbations with respect to the redshift for different models \ref{tab:alpha}. The top and bottom panels are for baryons and UDF respectively. Where the orange, dashed, dotted and green curved lines are for $\alpha=0,10^{-3},10^{-2},10^{-1}$ respectively. The horizon line denotes the limit of linear perturbation, i.e. $\delta=1$. The vertical parts of the curved lines denote the collapse of the perturbed regions. The corresponding turnaround redshifts can be found in the Table \ref{tab:alpha}.}\label{fig:per1}
\end{figure}
\end{center}

\begin{center}
\begin{figure}[htb]
\includegraphics[width=8cm]{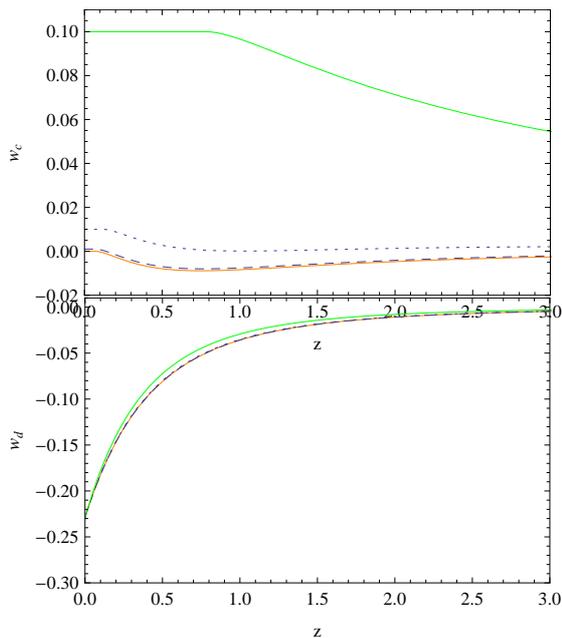}
\caption{The evolutions of $w_c$ and $w_d$ with respect to the redshift $z$ for different models \ref{tab:alpha}. The top and bottom panels are for $w_c$ and $w_d$ respectively. Where the orange, dashed, dotted and green curved lines are for $\alpha=0,10^{-3},10^{-2},10^{-1}$ respectively. Larger values of $\alpha$ result in values of $w_c$ closer and larger to $0$ during the collapse.}\label{fig:ws1}
\end{figure}
\end{center}

To investigate the effects of the model parameter $\alpha$ to the spherical collapse of baryon and UDF, we fix the other relevant model parameter $B_s$ to its central value $B_s= 0.229$ and initial conditions $\delta_{d}(z=1000)=3.5\times 10^{-3}$ and $\delta_{b}(z=1000)=10^{-5}$ by varying the model parameter $\alpha$ in the range from $0$ to $1$. We show the calculated results in Table \ref{tab:alpha}, where the redshift $z_{ta}$ is defined as the turnaround redshift when the perturbed region begin to collapse \cite{ref:Fernandes2012}. The evolutions of the density perturbations for baryons and UDF are shown in Figure \ref{fig:per1}. The vertical parts of the curved lines denote the collapse of the perturbed regions. The corresponding turnaround redshifts can be found in the Table \ref{tab:alpha}. Higher values of $\alpha$ result in values of $w_c$ closer and higher to $0$ during the collapse as shown in Figure \ref{fig:ws1}. Then the perturbations collapse earlier for the larger values of $\alpha$ as shown in Figure \ref{fig:per1}.

\begin{table}[tbh]
\begin{center}
\begin{tabular}{cccc}
\hline\hline Model & $\alpha$ & $z_{ta}$ & $\delta_b(z_{ta})/\delta_d(z_{ta})$ \\ \hline
a & $0$ & $0.0678$ & $ 1.240$\\
b & $10^{-3}$ & $0.111$ & $1.211$\\
c & $10^{-2}$ & $0.138$ & $0.689$\\
d & $10^{-1}$ & $0.940$ & $0.010$ \\
\hline\hline
\end{tabular}
\caption{Models for the STHC model, where the values of $\alpha$ are small positive values due to the constraint from background evolution history. The redshift $z_{ta}$ denotes the turnaround redshift when the perturbed region begin to collapse.}\label{tab:alpha}
\end{center}
\end{table}

%\begin{widetext}
%\end{widetext}

%\begin{widetext}
%\end{widetext}

\begin{center}
\begin{figure}[htb]
\includegraphics[width=8cm]{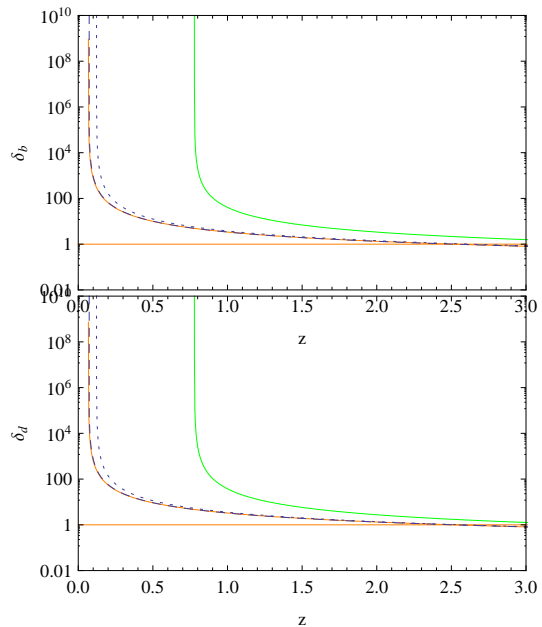}
\caption{The evolutions of density perturbations with respect to the redshift for different models \ref{tab:ics}. The top and bottom panels are for baryons and UDF respectively. Where the orange, dashed, dotted and green curved lines are for the models $\alpha=0,10^{-3},10^{-2},10^{-1}$ respectively. The horizon line denotes the limit of linear perturbation, i.e. $\delta=1$. The vertical parts of the curved lines denote the collapse of the perturbed regions. The corresponding turnaround redshifts can be found in the Table \ref{tab:ics}.}\label{fig:per2}
\end{figure}
\end{center}

It's time to show the influence of ICs on the evolution of the perturbations. Here we adopted different values of ICs for the UDF as shown in Table \ref{tab:ics}. The corresponding evolutions of density perturbations are shown in Figure \ref{fig:per2}. One can see that smaller value of initial conditions can result in later collapse as expected.  

\begin{table}[tbh]
\begin{center}
\begin{tabular}{ccccc}
\hline\hline Model & $\alpha$ & ICs ($\times 10^{-3}$) & $z_{ta}$ & $\delta_b(z_{ta})/\delta_d(z_{ta})$ \\ \hline
a & $0$ & $3.500$ & $0.0678$ & $1.240$\\
b & $10^{-3}$ & $3.490$ & $0.111$ &$1.211$ \\
c & $10^{-2}$ & $3.460$ & $0.122$ &$0.862$\\
d & $10^{-1}$ &  $3.210$ & $0.778$ & $0.00847$\\
\hline\hline
\end{tabular}
\caption{Models for the STHC model, where different initial values of UDF are adopted. The corresponding turnaround redshifts are summarized.}\label{tab:ics}
\end{center}
\end{table}

%\begin{widetext}
%\end{widetext}

\begin{center}
\begin{figure}[htb]
\includegraphics[width=8cm]{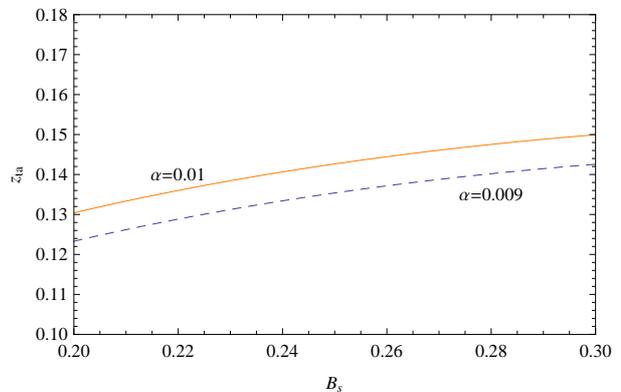}
\caption{The turnaround redshift $z_{ta}$ with respect to model parameter $\alpha$ for different values of $\alpha$. Where the orange solid line is for $\alpha=0.01$ and the dashed line is for $\alpha=0.009$. For plotting this figure, the initial conditions $\delta_{d}(z=1000)=3.5\times 10^{-3}$ and $\delta_{b}(z=1000)=10^{-5}$ are adopted.}\label{fig:zta}
\end{figure}
\end{center}

In the above, we have fixed the values of model parameter $B_s$ to its central value obtained by cosmic background evolution constraint. We should also investigate the effects to the collapse of the UDF of this model parameter. As pointed in our previous work \cite{ref:darkdeneracyxu}, the model parameter $B_s$ effectively takes the role as the ratio of cold dark matter in our Universe. Then it is easy to understand that the larger values of $B_s$ will increase the turnaround redshift $z_{ta}$. To see that, we show the $z_{ta}$ with respect to $B_s$ for different values of $\alpha$ in Figure \ref{fig:zta}.

%\begin{widetext}
%\end{widetext}

Through the above calculation and analysis, one can clearly see the possible large scale structure formation in this CASS model and understand the effects of the model parameters to the evolutions of the density perturbations.

\section{Conclusion} \label{sec:conclusion}

In this paper, we checked the possibility of the large scale structure formation for a UDF model which has a constant adiabatic sound speed (CASS) in the framework of spherical collapse. By choosing the different values of model parameters $\alpha$ and $B_s$, we show their effects on the non-linear perturbation evolutions. The results show that larger values of $\alpha$ and $B_s$ can make the density perturbations collapse earlier. After the investigation, we can conclude that it is possible to form large scale structure in the CASS UDF model via the spherical collapse. And the next step, we should study the hydrodynamical/$N$-body numerical simulation and compare the simulated results with the observed large scale strutter of our Universe. 

\section{Acknowledgements} L. Xu's work is supported in part by NSFC under the Grants No. 11275035 and "the Fundamental Research Funds for the Central Universities" under the Grants No. DUT13LK01.

\end{document}